\documentclass[aps,prl,twocolumn,superscriptaddress,showpacs]{revtex4}
\usepackage{graphicx}
\usepackage{times}
\begin{document}
\title{Electronic structures and edge effects of Ga$_2$S$_2$ nanoribbons}

\author{Bao-Ji Wang}
\affiliation{School of Physics and Chemistry, Henan Polytechnic University,2001 Shiji Road, Jiaozuo 454000, P. R. China}

\author{Xiao-Hua Li}
\affiliation{School of Physics and Chemistry, Henan Polytechnic University,2001 Shiji Road, Jiaozuo 454000, P. R. China}

\author{Li-Wei Zhang}
\affiliation{School of Physics and Chemistry, Henan Polytechnic University,2001 Shiji Road, Jiaozuo 454000, P. R. China}

\author{Guo-Dong Wang}
\affiliation{School of Electrical Engineering £¦ Automation, Henan Polytechnic University,2001 Shiji Road, Jiaozuo 454000, P. R. China}

\author{San-Huang Ke}
\email[Corresponding author, E-mail: ]{shke@tongji.edu.cn}
\affiliation{MOE Key Labortoray of Microstructured Materials, School of
Physics Science and Engineering, Tonji University, 1239 Siping Road, Shanghai 200092, P. R. China}
\affiliation{Beijing Computational Science Research Center, 10 Dongbeiwang West
Road, Beijing 100094, P. R. China}

%\date{\today{}}

\begin{abstract}
\textit{Ab initio} density functional theory calculations are carried out to predict
the electronic properties and relative stability of
gallium sulfide nanoribbons (Ga$_2$S$_2$-NRs)
with either zigzag- or armchair-terminated edges. It is found that the
electronic properties of the nanoribbons are very sensitive to the edge structure.
The zigzag nanoribbons (Ga$_2$S$_2$-ZNRs) are metallic with spin-polarized edge
states regardless of the H-passivation,
whereas the bare armchair ones (Ga$_2$S$_2$-ANRs) are semiconducting with an
indirect band gap. This band gap exhibits an oscillation behavior as the width
increases and finally converges to a constant value.
Similar behavior is also found in H-saturated Ga$_2$S$_2$-ANRs,
although the band gap converges to a larger value.
The relative stabilities of the bare ANRs and ZNRs are investigated
by calculating their binding energies. It is found that for a similar width the
ANRs are more stable than the ZNRs, and both are more stable than
some Ga$_2$S$_2$ nanocluters with stable configurations.

\end{abstract}
\maketitle

\section{Introduction}

Low-dimensional materials, especially one-dimensional ($1$-D) nanoribbons,
have attracted significant attention from the scientific community
during the past two decades due to their interesting electronic
properties associated with their low dimensionality and the
resulting quantum confinement effect. In the past few years,
graphene nanoribbons (GNRs) -- thin strips of graphene -- have been extensively
studied
because of their rich and exotic physical properties which depends on 
their size and edge termination\cite{Nakada1996-17954,zhao2009-3012}.
First-principles calculations have revealed that GNRs with hydrogen saturated
armchair- or zigzag-shaped edges always have a nonzero direct band gap
which decreases as the ribbon width increases. The band gap variation
for armchair GNRs (AGNRs) exhibits distinct "family behaviors"\cite{Son2006-216803}.
The band gaps of AGNRs arise from both quantum confinement
and the crucial effect of the edges, while for zigzag GNRs (ZGNRs), gaps appear
because of a staggered sublattice potential on the hexagonal lattice
due to edge magnetization\cite{Son2006-216803,Okada2001-87,Son2006-444,Nakada1996-17954}.
Besides GNRs, some other layer-structured nanoribbons,
such as boron nitride nanoribbons (BNNRs) and MoS$_2$ nanoribbons, have also been studied intensively\cite{Xu2013-3766}.
It is found that the band gap of hydrogen-terminated zigzag BNNRs is indirect
and decreases monotonically with the increasing ribbon width. Whereas, direct band gap
oscillation is observed for armchair BNNRs\cite{Nakamura2005-205429,Du2007-181},
which tends to converge to a constant value when the ribbon is wider than 3 nm.
In contrast, zigzag edged MoS$_2$ nanoribbons show a metallic behavior irrespective 
of the ribbon width, while armchair edged ones are semiconducting and the band gaps converge to
a constant value as the ribbon width increases\cite{Li2008-16739}.

As a wide indirect-band-gap semiconductor with uniform layered structure,
gallium sulfide (GaS) \cite{Ho2006-083508} has been used in photoelectric devices,
electrical sensors, and nonlinear optical applications and gained renewed interest
\cite{Shen2009-1115}. In its bulk form, GaS usually crystallizes into a
layered structure, S-Ga-Ga-S (Ga$_2$S$_2$), in which each layer consists of two
AA-stacked hexagonal sublayers of Ga atoms sandwiched
between two hexagonal sublayers of S atoms. These layers are bound
in a three-dimensional ($3$-D) structure by the nonbonding interaction through the S atoms
along the vertical axis\cite{Shen2009-1115}. Recently, the micromechanical
cleavage technique (as originally used in peeling off graphene from graphite)
was successfully used to fabricate single-layer sheets of Ga$_2$S$_2$ \cite{Late2012-3549,Tang2013-1244}.
A theoretical calculation based on the density functional theory (DFT) has showed that
the Ga$_2$S$_2$ sheet is dynamically stable and is a
indirect-band-gap semiconductor with an unusual inverted
sombrero dispersion of holes near the top of the valence band\cite{lyomi2013-195403}.
As for the electronic properties of Ga$_2$S$_2$ nanoribbons (Ga$_2$S$_2$-NRs) which are important
in realistic device applications,
very few reports are available in the literature, to the best of our knowledge.

The purpose of this work is to investigate the electronic structures
of Ga$_2$S$_2$-NRs by performing
first-principles DFT calculations. The effects of different ribbon widths and
different edge structures are studied. Also studied are the relative stabilities
of these nanoribbons.
In the next section, we describe our computational method.
Section III presents our calculated results, and the last section is devoted to conclusions.

\section{Structure and Computation}

The models of Ga$_2$S$_2$-NRs are constructed by cutting out a stripe of Ga$_2$S$_2$
sheet with the desired edges and widths. Adopting a similar notation used to describe
GNRs\cite{Nakada1996-17954,zhao2009-3012,Wakabayashi1999-8271},
we use the number of zigzag lines (\textit{N}$_z$) or dimer lines (\textit{N}$_a$)
to present the width of a zigzag-edged ribbon (Ga$_2$S$_2$-ZNR) or a
armchair-edged ribbon (Ga$_2$S$_2$-ANR) and denote the ribbon by
\textit{N}$_z$-Ga$_2$S$_2$-ZNR (as illustrated in Fig.~\ref{fig:structure-z-a}(a))
or \textit{N}$_a$-Ga$_2$S$_2$-ANR
(as illustrated in the Fig.~\ref{fig:structure-z-a}(b)).

Calculations of structure optimization and band structure are carried out by
using a first-principles pseudopotential plane wave method based on DFT, as
implemented in the Vienna {\it ab initio} simulation package (VASP)\cite{Kresse1993-558}.
The projector augmented wave method (PAW)\cite{Blochl1994-17953} is used to describe
the ion-electron interaction. The electron exchange and correlation are treated
by the generalized gradient approximation (GGA)\cite{Perdew1992-6671} in the version of Perdew-Burke-Ernzerhof
(PBE)\cite{Perdew1996-3865}. A kinetic energy cutoff of $600$\,eV is adopted for
the plane-wave expansion of the wave function.
The $4s^{2}4p^{2}$ electrons of Ga, and $3s^{2}3p^{4}$ electrons of S are
treated as valence electrons.
Periodic boundary conditions (PBC) are employed for the infinitely long
nanoribbon systems and a vacuum space of $10$\,\AA\, in each direction
perpendicular to the ribbon is used to eliminate the interaction between
the periodic images.
The atomic structures of the systems, including the lattice parameters and
atomic positions, are fully relaxed by the conjugate gradient method until the residual forces
acting on each atom are less than $0.001$\,eV/\AA\,.
For the structure optimization a Monkhorst-Packs k-mesh of $1 \!\times\! 1
\!\times\! 12$ is used for the Brillouin zone (BZ) sampling.
The band structures are presented by $45$ k-points along the $X$-axis in the BZ.

\section{Results and Discussion}

\begin{figure}[htbp]
\includegraphics[width=3.0 in,clip]{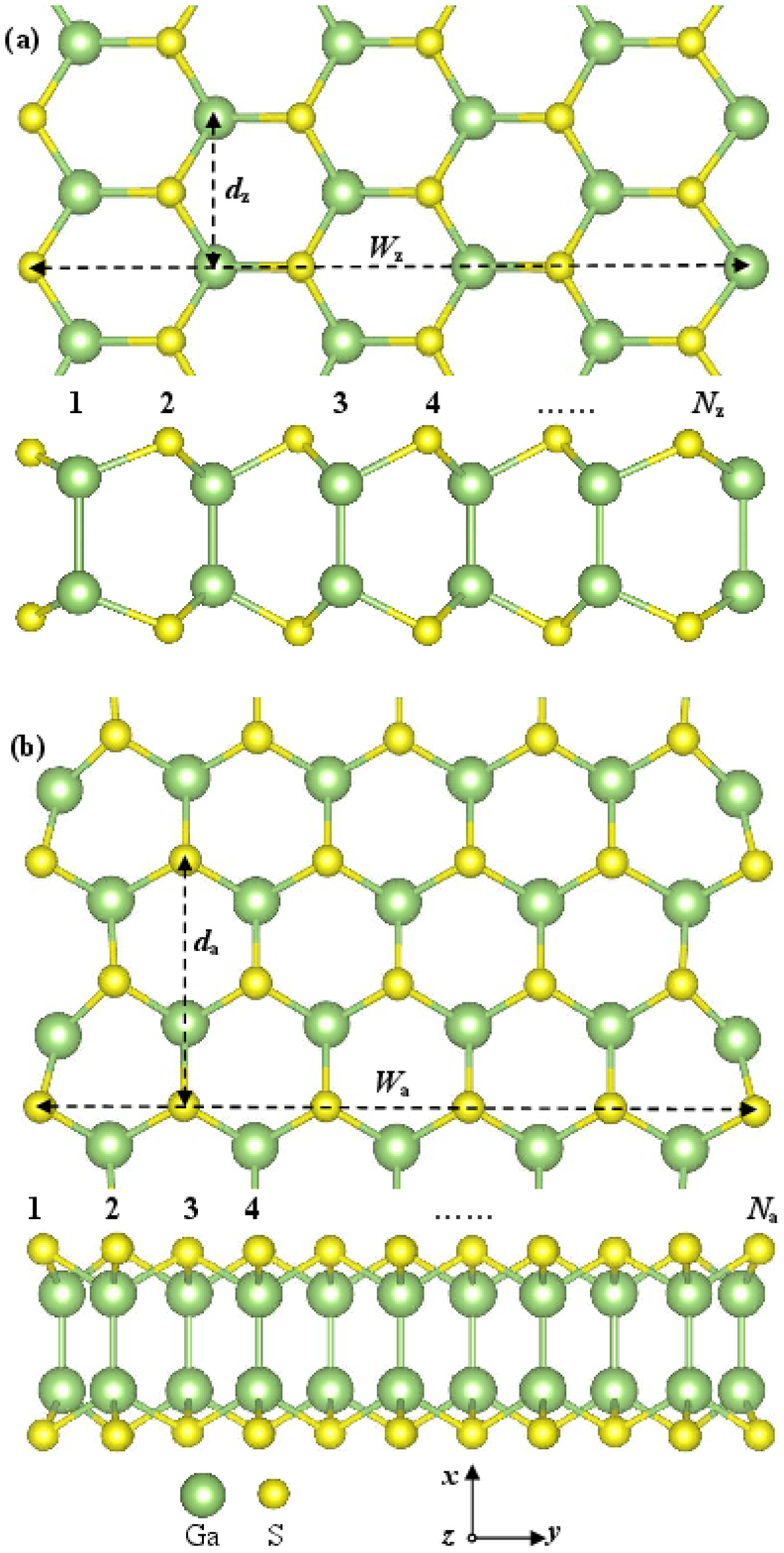}
\caption{\label{fig:structure-z-a} Top and side views of the atomic structures of
$6$-Ga$_2$S$_2$-ZNR (a) and $11$-Ga$_2$S$_2$-ANR (b). The ribbon width
and $1$-D unit cell distance are denoted by $W_{z}$ ($W_{a}$)
and $d_{z}$ ($d_{a}$) for the ZNRs (ANRs), respectively. The ribbons are extended periodically
along the $x$ direction.}
\end{figure}

\begin{figure*}[htbp]
\includegraphics[width=6.0 in,clip]{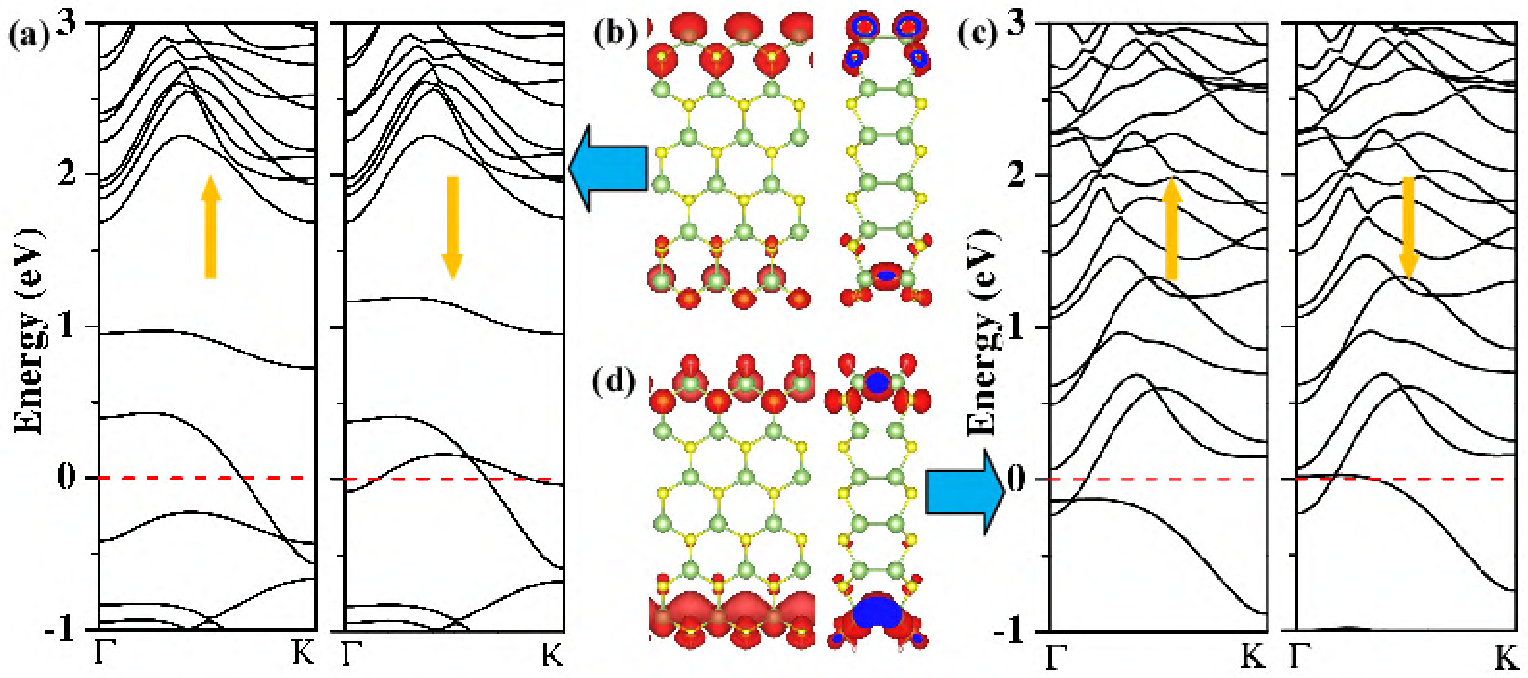}
\caption{\label{fig:6-zigzag-band-densities} Calculated spin-polarized band structures
(left: majority spin; right: minority spin) and corresponding local density of
states (LDOS) near the Fermi level for bare $6$-Ga$_2$S$_2$-ZNR ((a),(b)) and H-terminated $6$-Ga$_2$S$_2$-ZNR ((c),(d)).
}
\end{figure*}

\begin{figure*}[htbp]
\includegraphics[width=6.0 in,clip]{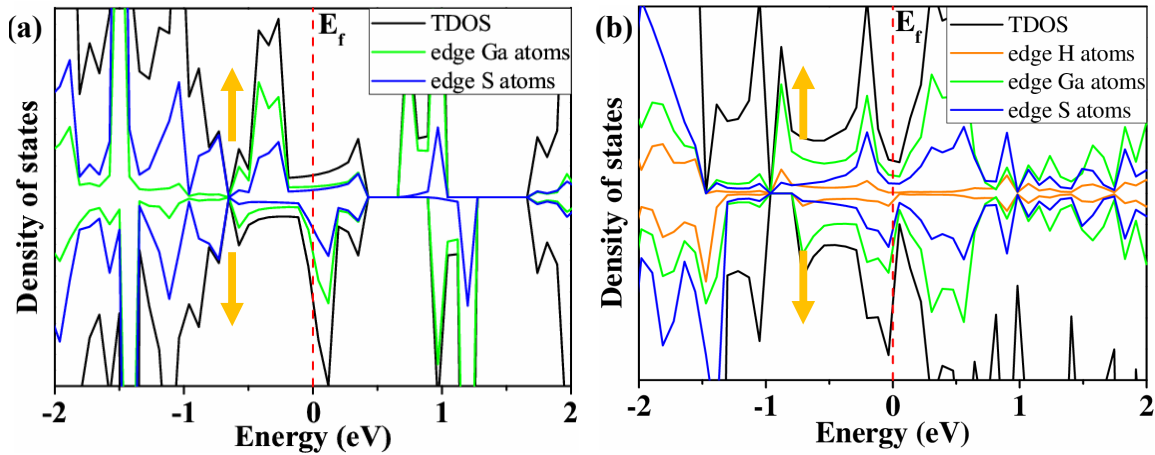}
\caption{\label{fig:6-zigzag-dos} Spin-dependent total and partial density of states
for bare $6$-Ga$_2$S$_2$-ZNR (a) and H-terminated $6$-Ga$_2$S$_2$-ZNR (b).
}
\end{figure*}

We first optimize the atomic structure of an periodic 2-D Ga$_2$S$_2$ sheet,
in which Ga-S and Ga-Ga atoms are linked covalently in the quadruple layer.
The optimized bond lengths of Ga-S and Ga-Ga are $2.37$\,\AA\,and $2.47$\,\AA, respectively.
The Ga-S-Ga (or S-Ga-S) and Ga-Ga-S bond angles are $100.38^{\circ}$ and $117.50^{\circ}$,
respectively. The thickness of the quadruple layer is $4.76$\,\AA.
The computed binding energy is $3.64$\,eV per atom. The Ga$_2$S$_2$ sheet is
found to be a semiconductor with an indirect band gap of $2.38$\,eV on
the DFT/PBE level. These results are in good agreement with recent DFT
calculations\cite{lyomi2013-195403,Kohler2004-193403}.
We note that Kohn-Sham DFT calculations usually underestimate band gaps of semiconductors
because of the lacking of the derivative discontinuity in the energy functionals.
A hybrid DFT calculation\cite{lyomi2013-195403} gives a larger value of the
indirect band gap, $3.28$\,eV.
Although the Kohn-Sham DFT has the band gap issue, it can usually give reasonable
shapes of energy bands and wave functions in the vicinity of the Fermi level,
compared to results from quasiparticle GW calculations.

\begin{figure*}[htbp]
\includegraphics[width=6.0 in,clip]{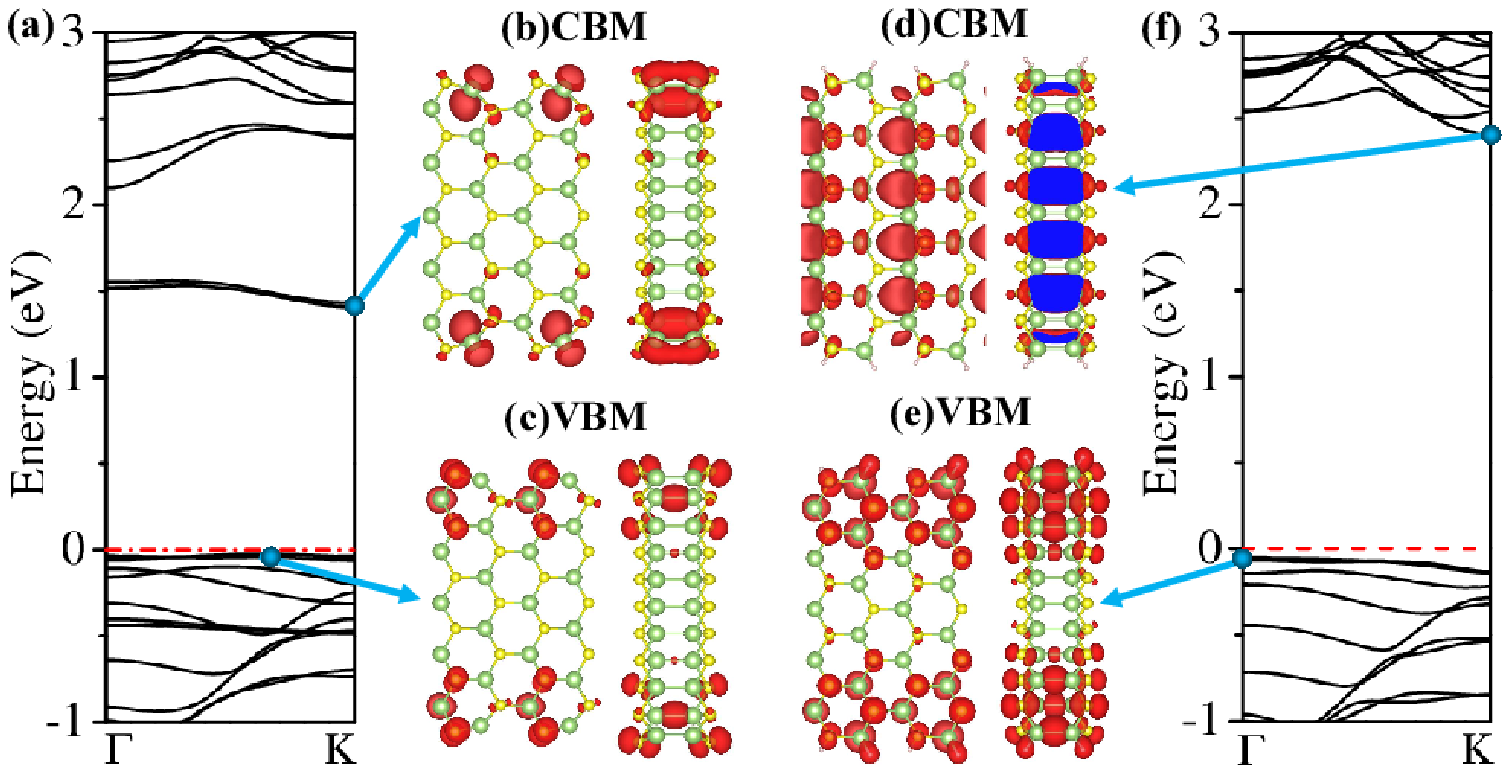}
\caption{\label{fig:11-armchair-band-densities} Band structures and
corresponding LDOS  
for CBM and VBM of the $11$-Ga$_2$S$_2$-ANR. (a),(b),(c) correspond to the bare
Ga$_2$S$_2$-ANR, (d), (e), (f) correspond to the H-terminated Ga$_2$S$_2$-ANRs.
Each LDOS figure contains
a top view (left) and a side view (right) of the LDOS isosurface of
the specified states at CBM or VBM. The Fermi level is set to be zero.
}
\end{figure*}

\begin{figure}[htbp]
\includegraphics[width=3.0 in,clip]{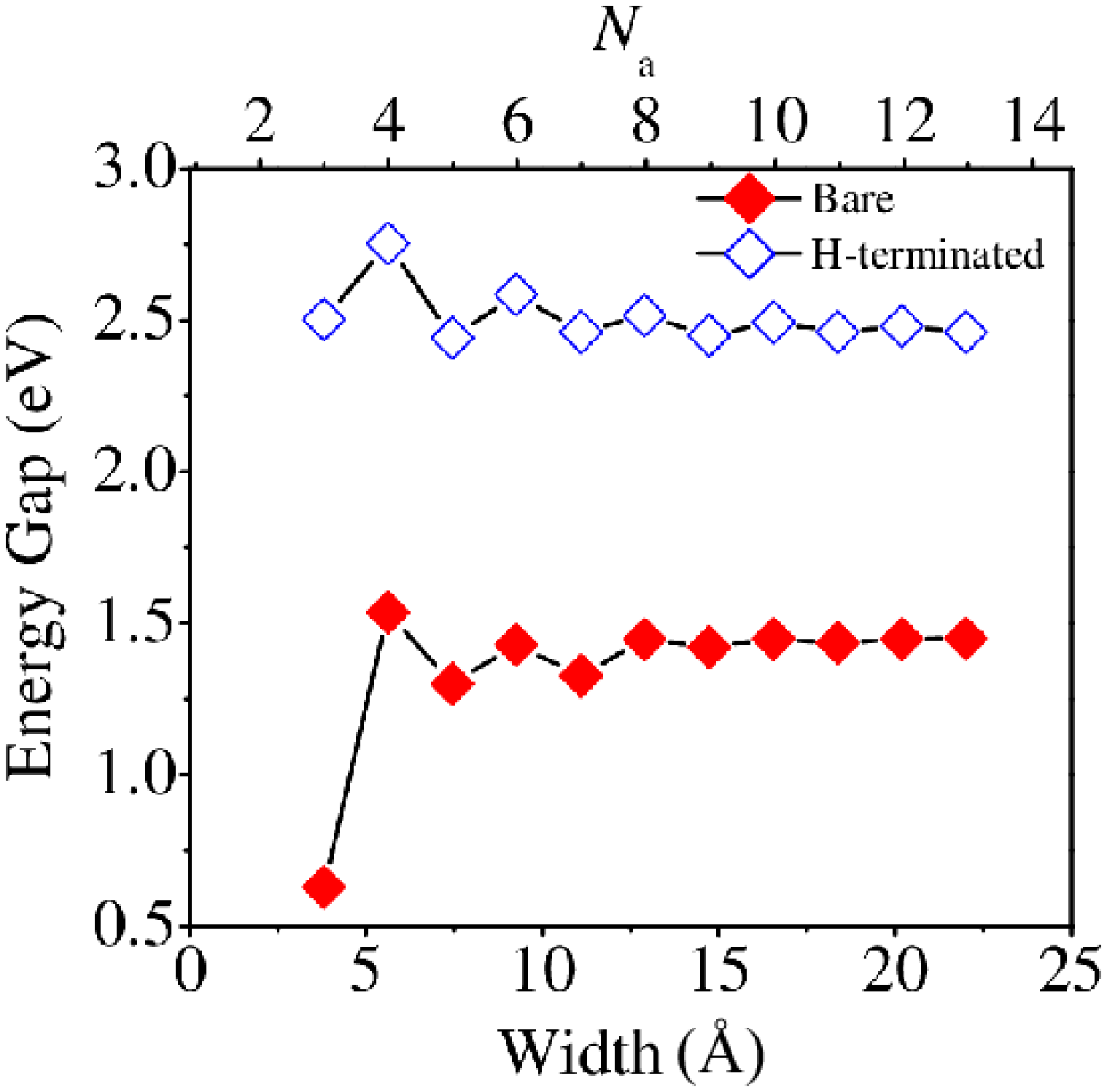}
\caption{\label{fig:bandgap-w-n} Band gaps of a series of bare
Ga$_2$S$_2$-ANRs ($3 \leq N_{a}\leq 13$) as a function of the ribbon widths
(filled red lozenges). For a comparison, the result for the corresponding hydrogen saturated
Ga$_2$S$_2$-ANRs is also shown (voided blue lozenges).
}
\end{figure}

We consider two different directions of termination for Ga$_2$S$_2$-NRs, zigzag
and armchair, and for each direction of termination we consider the bare and
H-saturated edge configurations.
In order to investigate the width dependence of the electronic properties,
we perform calculations for different ribbon widths, up to $N_z$=24 and
$N_a$=19 (i.e., $24$-Ga$_2$S$_2$-ZNR and $19$-Ga$_2$S$_2$-ANR).
Since qualitatively similar results are observed for the different widths,
here we first take the $6$-Ga$_2$S$_2$-ZNR with width of $17.37$\,\AA\,
and $11$-Ga$_2$S$_2$-ANR with width of $18.37$\,\AA\,
as prototypes to present our results (see Fig.~\ref{fig:structure-z-a}).
After the structure optimization, the bond length between Ga and S atoms
varies depending on their position in the ribbon.
For the $6$-Ga$_2$S$_2$-ZNR, the Ga-S bond lengths are $2.36$\,\AA\,and $2.37$\,\AA\,
at inner sites, and $2.41$\,\AA\,and $2.24$\,\AA\, at the Ga-terminated
and S-terminated edges, respectively. The bond angles of Ga-Ga-S
and S-Ga-S at the inner sites are $117.43^{\circ}$ and $100.24^{\circ}$, respectively.
Because of the edge relaxation, the two angles change to $105.60^{\circ}$ 
at the Ga-terminated edge and $108.48^{\circ}$ at the S-terminated edge. 
In the case of the $11$-Ga$_2$S$_2$-ANR, the edge S atoms tend to shift outward slightly. 
The Ga-S bond lengths are $2.36$\,\AA, $2.37$\,\AA\,and $2.41$\,\AA\, at the inner sites
and $2.21$\,\AA\, at the two edges. The Ga-Ga bond lengths are $2.48$\,\AA, $2.47$\,\AA\,
and $2.50$\,\AA\, at the inner sites and $2.48$\,\AA\, at the two edges.
The bond angles of Ga-Ga-S and S-Ga-S are $117.50^{\circ}$ and $100.36^{\circ}$,
respectively, at the inner sites. These results show strong edge-induced atomic
relaxation which results in quite different bond lengths and angles at the edges
compared to those in $2$-D Ga$_2$S$_2$ sheet. This behavior is quite different 
from the cases of GNRs and BNNRs where the edge distortion is much smaller.

Our calculations for the ZNRs show that their electronic structures are spin
polarized. As an example, the band structures of $6$-Ga$_2$S$_2$-ZNR with bare and
H-saturated edges are shown in Figs.~\ref{fig:6-zigzag-band-densities}(a),(c) for the
two spin components: up-spin (majority) and down-spin (minority), respectively.
One can see that the band structures of $6$-Ga$_2$S$_2$-ZNR exhibit a metallic nature: 
There are one or two bands crossing the Fermi level regardless of the H-passivation, 
which is similar to the behavior of MoS$_2$ Nanoribbons\cite{Ataca2011-3934,Pan2012-7280}.
We find that this metallic nature holds for all Ga$_2$S$_2$-ZNRs studied
(up to $N_z$= 24) regardless of the ribbon width.
One should note that the difference between the up-spin and down-spin band
structures exists only around the Fermi energy, indicating that the spin
polarization originates from the states near the Fermi energy. 

To understand the characteristics of these bands near the Fermi energy, we plot 
their local density of states (LDOS) in
Fig.~\ref{fig:6-zigzag-band-densities}(b),(d) for the bare and H-saturated
ribbons, respectively.
It can be seen that in both cases the LDOS is located around the edges,
indicating that the metallic nature of Ga$_2$S$_2$-ZNRs is mainly due to the
spin-polarized edge states.
To futher show the origin of these metallic bands, we also calculate the total density
of states (TDOS) and projected density of states (PDOS) of $6$-Ga$_2$S$_2$-ZNR (see Fig.~\ref{fig:6-zigzag-dos}).
The states around the Fermi level in both spin channels are dominated
by the $4s$ and $4p$ states of edge Ga atoms and $3p$ states of edge S atoms.
Further Mulliken population analyses show that the spin polarization is mainly due
to the ferromagnetic state of the Ga-terminated edge regardless of the 
H-saturation.

In contrast, our calculations for the ANRs show that their electronic structures
are spin unpolarized. Both the bare and H-saturated Ga$_2$S$_2$-ANRs 
exhibit a semiconducting nature.
As an example shown in Fig.~\ref{fig:11-armchair-band-densities}(a), 
bare $11$-ZGa$_2$S$_2$-ANR is a typical semiconductor with an indirect band gap of $1.44$\,eV.
In this case, the valence-band maximum (VBM) occurs along the $\Gamma-K$ direction,
and the conduction band minimum (CBM) lies at the $K$ point.
We note that the lowest conduction band and highest valence band are very flat, which is
similar to the case of BNNRs\cite{Park2008-2200}.
When the edge Ga and S atoms are saturated by H atoms,
Ga$_2$S$_2$-ANRs are still indirect band gap semiconductors but with a lager
band gap (see Fig.~\ref{fig:11-armchair-band-densities}(f)) and the VBM and CBM
are now located at the $\Gamma$ and $K$ point, respectively.

To analyze the characteristics of the VBM and CBM and the influence of the edges, we calculate
the LDOS isosurfaces for the energies around the VBM and CBM of the bare
and H-saturated 11-Ga$_2$S$_2$-ANR (see
Fig.~\ref{fig:11-armchair-band-densities}(b),(c),(d), and (e)). 
In the bare system the large edge relaxation (reconstruction) may weaken the
dangling-bond states. However, the states of VBM and CBM are still mainly determined
by the dangling-bond-related states which are driven from Ga-$4p$ and S-$3p$
orbitals. Therefore, their LDOS are mainly localized around the edges of the ribbon (see
Fig.~\ref{fig:11-armchair-band-densities}(b),(c)) and form flat bands\cite{Ataca2011-3934}.
Because of this edge-state effect, the converged band gap ($1.45$\,eV) (see the discussion later) of bare
Ga$_2$S$_2$-ANRs becomes smaller than that of the $2$-D Ga$_2$S$_2$ sheet ($2.38$\,eV).

When the edge Ga and S atoms are saturated by H atoms,
the dangling-bond edge states disappear and thus the band gap increases.
As can be seen in Fig.~\ref{fig:11-armchair-band-densities} (d),
now the LDOS of the CBM are mainly located near the center of the ribbon.
For the VBM, its LDOS are nearly uniformly distributed throughout the ribbon
though very weak edge states are remained (see
Fig.~\ref{fig:11-armchair-band-densities}(e)).
In contrast with the bare systems, now the converged band gap (2.45eV) (see the discussion later) of H-saturated
Ga$_2$S$_2$-ANRs becomes slightly larger than that of the $2$-D Ga$_2$S$_2$ (2.38eV).

Next let us look at the variation of band gap as a function of ribbon width for both the bare and H-sturated ANRs.
As shown in Fig.~\ref{fig:bandgap-w-n}, all the Ga$_2$S$_2$-ANRs are semiconducting
and their band gaps are sensitive to the structural symmetry and
ribbon width. Generally, owing to the quantum confinement effect the band gap $E_G$
should decrease with increasing ribbon width \textit{N}$_a$. However, here
the variation of $E_{G}$ with \textit{N}$_a$ shows an oscillatory (or family)
behavior, especially for the narrow ribbons (\textit{N}$_{a} < 8$).
Specifically, antisymmetric Ga$_2$S$_2$-ANRs
(\textit{N}$_{a} \!=\! 2m$, where $m$ is an integer) have larger band gaps
than the neighboring symmetric ones (\textit{N}$_{a}\!=\!2m-1$).
This 'family behavior' was also observed previously in armchair-edged GNRs\cite{Son2006-216803},
BNNRs\cite{Park2008-2200}, and MoS$_2$ nanoribbons \cite{Li2008-16739,Cai2014-6269},
where a width-dependent oscillation in band gap occurs.
When the ribbon becomes wide enough ($>$ $13$\,\AA\, or \textit{N}$_{a}\!>\!8$),
the band gaps tend to converge to a constant value ($\sim 1.45$\,eV),
being smaller than the bulk band gap of the $2$-D Ga$_2$S$_2$ sheet ($2.38$\,eV).
To carefully confirm this tendency, we further do a calculation for
$19$-Ga$_2$S$_2$-ANRs (width $33$\,\AA), which also shows a band gap of $1.45$\,eV.
In the case of H-terminated ANRs (see Fig.~\ref{fig:bandgap-w-n}),
similar behavior is found although their band gaps converge to a larger value of
$2.46$\,eV due to the diminishing edge states.
The variation of the band gap with \textit{N}$_a$ may be an important advantage for their
applications in nanotechnology because it may lead to formation of quantum dot or
multiple quantum wells through the width modulation\cite{Sevincli2008-245402,Li2008-16739}.

\begin{figure}[htpb]
\includegraphics[width=3.0 in,clip]{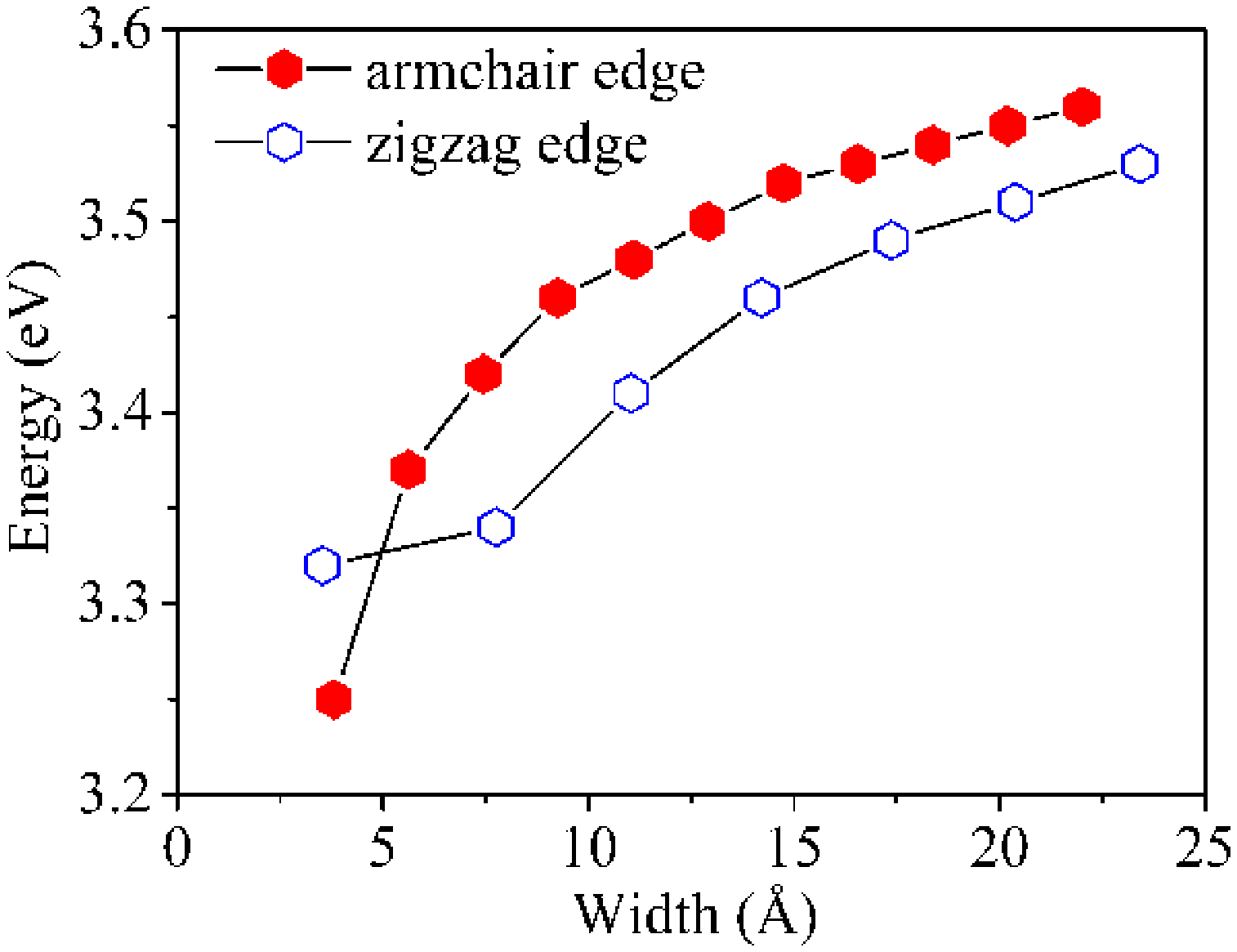}
\caption{\label{fig:bindenergy-w} Binding energy of bare Ga$_2$S$_2$-ZNRs ($3\leq N_{z}\leq 8$) and Ga$_2$S$_2$-ANRs ($6 \leq N_{a}\leq 13$) as a function of the ribbon width.}
\end{figure}

Finally, we would like to discuss the stability of Ga$_2$S$_2$-NRs,
which is quite important because it determines
whether this nanostructure can be realized experimentally.
The stabilities of different nanoribbons and nanoclusters can be evaluated with
their binding energies; those with larger binding energies would be more stable. 
To estimate the stability, we calculate the binding energy per atom as a function of 
the ribbon width for bare Ga$_2$S$_2$-NRs in the following way\cite{Li2008-16739}:
$E_{b} \!=\!(nE_{\text{Ga}} + mE_{\text{S}}-E_{\text{Ga}_{n}\text{S}_{m}})$/($n + m$)
where \textit{E}$_{\text{Ga}}$, \textit{E}$_{\text{S}}$,
and $E_{\text{Ga}_{n}\text{S}_{m}}$ are the total energies of Ga, S atoms, and
Ga$_n$S$_m$, respectively, with $n$ and $m$ being the number of Ga and S atoms, respectively.
As is shown in Fig.~\ref{fig:bindenergy-w}, the binding energy increases monotonically
with increasing ribbon width for both zigzag and armchair Ga$_2$S$_2$-NRs.
The binding energies of the armchair ribbons are slightly higher than those of
the zigzag ribbons with comparable widths, indicating that armchair
Ga$_2$S$_2$-NRs are more stable than zigzag ones. It is important to note
that although the binding energies of the nanoribbons (except for
\textit{N}$_{a}\!=\!$\,3) is smaller than the 2-D Ga$_2$S$_2$ sheet,
it is larger than that of Ga$_2$S$_2$ nanocluters with stable
configurations\cite{Dwivedi12012-68}.
Therefore, it is reasonable to expect that stable bare Ga$_2$S$_2$-NRs are
possible to make experimentally in the future. 
Since dangling bonds are saturated by H atoms in H-saturated Ga$_2$S$_2$-NRs,
their stability should be even better.

\section{Conclution}

In summary, by performing first-princeples DFT calculations we have investigated
systematically the electronic properties
of Ga$_2$S$_2$ nanoribbons which have not been
realized yet but the corresponding 2-D sheet has been successfully fabricated experimentally.
We have considered the bare and H-saturated zigzag- and armchair-edged
nanoribbons with different ribbon widths (up to 3 nm).
It is found that the electronic properties of the nanoribbons strongly
depends on their edge structures. The Ga$_2$S$_2$-ZNRs are
metallic with spin-polarized edge states while the Ga$_2$S$_2$-ANRs are indirect-band-gap
semiconductors without spin polarization.
The band gap of bare Ga$_2$S$_2$-ANRs exhibits an oscillation behavior with increasing ribbon
width and ultimately converges to a constant value of $1.45$\,eV when the
ribbon becomes wide enough ($>$ 13 {\AA}). The H-saturated Ga$_2$S$_2$-ANRs have
the similar behavior but with larger converged band gap of $2.46$\,eV due to the
saturation of the dangling bonds at the edges.
Finally, we have studied the relative stabilities of the bare ZNRs and ANRs by
calculating their binding energy per atom, and
found that the ANRs are more stable than the ZNRs with a similar width. The
binding energies are found to be slightly smaller than that of the 2-D sheet,
but is larger than that of Ga$_2$S$_2$ nanocluters with stable configurations,
indicating that the bare ribbons can be expected to be realized experimentally
in the future.

\section {Acknowledgments}
This work was supported by the National Natural Science Foundation of
China (No. 11174220 and 11374226), and by the Key Scientific Research
Project of the Henan Institutions of Higher Learning (No. 16A140009),
and by the Program for Innovative Research Team of Henan Polytechnic
University (No. T2015-3), as well as by the Doctoral Foundation of
Henan Polytechnic University (No. B2015-46).

%\bibliography{Ga2S2}

\end{document}